\documentclass[twocolumn,showpacs,superscriptaddress,aps,prb]{revtex4}

\usepackage[dvips]{graphicx}
\usepackage{amssymb,amsmath}
\usepackage{dcolumn}
\usepackage{bm}
\newcommand{\bsigma}{\mbox{\boldmath$\sigma$}}

\newcommand{\btau}{\mbox{\boldmath$\tau$}}

\newcommand{\beq}{\begin{equation}}
\newcommand{\eeq}{\end{equation}}

\begin{document}
\title{Spin-polarized transport through domain wall in magnetized graphene}


\author{M. Khodas}
\email{mkhodas@bnl.gov}
\affiliation{Department of Physics, Brookhaven National Laboratory, Upton, NY 11973-5000}
\affiliation{Department of Condensed Matter Physics and Materials Science Brookhaven National Laboratory, Upton, NY 11973-5000}
\author{I. A. Zaliznyak}
\affiliation{Department of Condensed Matter Physics and Materials Science Brookhaven National Laboratory, Upton, NY 11973-5000}
\author{D. E. Kharzeev}
\affiliation{Department of Physics, Brookhaven National Laboratory, Upton, NY 11973-5000}
%

\date{\today}




\begin{abstract}
Atomically thin two-dimensional layer of honeycomb crystalline carbon known as graphene is a promising system for electronics. It has a point-like Fermi surface, which is very sensitive to external potentials. In particular, Zeeman magnetic field parallel to the graphene layer splits electron bands and creates fully spin-polarized and geometrically congruent circular Fermi surfaces of particle and hole type.
In the presence of electric field, particles and holes with opposite spins drift in opposite direction. These phenomena are likely to be of interest for developing graphene-based spintronic devices. A domain wall (DW) separating regions with opposite spin polarizations is a basic element of such a device. Here we consider a ballistic passage of spin-polarized charge carriers through DW in graphene. We also discuss the analogy between the generation of spin currents in graphene and in relativistic quark-gluon plasma, where the spin-polarized current is responsible for the phenomenon of charge separation studied recently at RHIC.
\end{abstract}
%


%
%
\maketitle

\section{Introduction}

The remarkable properties of graphene \cite{GeimNovoselov_NatMat2007} at present attract attention of many researchers.
Its honeycomb two-dimensional (2D) crystalline order is extremely robust.  In view of the well-known Peierls-Landau argument proving the thermodynamical instability of isolated 2D crystals \cite{LandauLifshitz}, robustness of graphene monolayers may look somewhat surprising. While graphene sheets in natural and synthetic graphite materials are not isolated but are supported on 3D substrates \cite{OshimaNagashima_JPCM1997,Forbeaux_PRB1998}, binding of the monolayers is so weak that they can be easily exfoliated and in many cases appear approximately isolated from the substrate \cite{Forbeaux_PRB1998,Novoselov_PNAS2005,Charrier_JAP2002,Ohta_PRL2007}.

Adding to the enthusiasm were recent discoveries of exceptionally high electronic quality of graphene monolayers \cite{Novoselov_Science2004,Berger_Science2006}. Carefully prepared samples show ambipolar electric field effect with carrier mobilities exceeding $ 10^4$ cm$^2$/V/s for electron/hole concentrations up to $\sim 10^{13}$ cm$^{-2}\ $. \cite{Novoselov_Science2004,Berger_Science2006}
High quality graphene samples at present can be
obtained either by using graphite exfoliation, which results in graphene pieces with 1 to 100 $\mu$m linear dimensions \cite{GeimNovoselov_NatMat2007}, or by epitaxial growth on SiC via silicon sublimation, which yields macroscopic mosaic layer with micron-size crystalline domains \cite{Hass_APL2006}.

High electron mobility implies ballistic charge transport and electronic phase coherence on the micron length scale, which is comparable with high-quality semiconductor heterostructures traditionally used for the studies of the quantum Hall effect (QHE) \cite{Tsui_etal_PRL1982}. Moreover, charge mobilities in graphene are only weakly temperature dependent, being probably limited by sample imperfections and size effects even at room temperature \cite{GeimNovoselov_NatMat2007}. Hence, not only QHE was indeed observed in graphene \cite{Novoselov_Nature2005,Zhang_Nature2005}, but it was also found to survive up to 300 K \cite{Novoselov_Science2007}, indicating that the electrons in graphene form a quantum gas even at room temperature.

\begin{figure}[h]
\begin{center}
\includegraphics[width=1.\columnwidth]{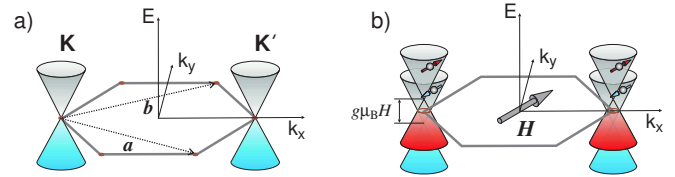}
\caption{(a) Electronic band structure resulting from the $sp^2$ C-C bonding in the hexagonal carbon layer of graphene. In zero magnetic field the filled $\pi$ and the empty $\pi^*$ bands meet at a single point, resulting in a linear 2D dispersion, $\varepsilon({\bf k}) = v_F \hbar k$, characteristic of 2D relativistic Dirac fermions ($v_F$ is Fermi velocity). {\bf a} and {\bf b} define triangular lattice of a honeycomb graphene layer containing two C atoms. Fermi ``surface'' consists of two inequivalent points $K$ and $K'$ (valleys). (b) Magnetic field $H$ parallel to graphene layer introduces Zeeman splitting $g\mu_B H$ between the bands with parallel (P) and antiparallel (AP) spin. P and AP bands acquire congruent Fermi-surfaces of hole- and electron-type respectively, whose radius is $\hbar k_F = g\mu_B H /(2 v_F)$.}
\label{f:bands}
\end{center}
\end{figure}

The exceptional electronic properties of graphene such as high
charge carrier mobility, long mean free path and coherence length and ability to support high current densities, exceeding $ 10^8$ A/cm$^2$, \cite{Novoselov_Science2004} make it a promising candidate for nano-scale electronics. In addition, its unique electronic structure provides a playground for the studies of (2+1)-dimensional (space+time) quantum electrodynamics. The crossing of the energy bands associated with two different sub-lattices $A$ and $B$ of the graphene's honeycomb crystal lattice results in the energy spectrum of electron and hole quasi-particles, which is linear in momentum $\hbar k$ [see Fig \ref{f:bands}(a)], $\epsilon(k) = v_F \hbar k$, where $v_F \simeq 10^6 \ {\rm m/s}$ is the Fermi velocity. This has been observed experimentally in transport measurements \cite{Novoselov_Nature2005,Zhang_Nature2005} and in angle-resolved photoemission \cite{Ohta_PRL2007}.

A gapless linear 2D spectrum of electron and hole quasi-particles belonging to two sublattices implies that charge carriers in graphene can be formally described as two-dimensional relativistic chiral fermions with spin and with pseudo-spin accounting for the two-sublattice band structure. A straightforward consequence is the conservation of the pseudo-spin chirality of the quasiparticles defined as the pseudo-spin projection on the momentum, $\mathbf{p} \cdot \btau $. The orbital motion in magnetotransport and QHE experiments  can be described by the ``truncated'' 2D Dirac equation for massless fermions \cite{Novoselov_Nature2005,Zhang_Nature2005,Novoselov_Science2007},
\begin{equation}
\label{DiracEq}
 ({\bf p} +\frac{e}{c}{\bf A}) {\btau} \psi_{\tau'} = \varepsilon \psi_{\tau'} \, ,
\end{equation}
where ${\btau_i} = (\tau^x,\tau^y)^T$ are Pauli matrices acting
in the pseudo-spin space. They account for the two-sublattice nature of graphene's honeycomb lattice and the resulting composite structure of the dispersion cone around each of the Fermi points. $\psi_{\tau'}$ is a rank two spinor wavefunction and $\bf p$ is the 2D momentum operator \cite{Khveshchenko,GorbarGusynin_PRB2002,DeMartino_PRL2007}.

Eq. (\ref{DiracEq}) assumes degeneracy with respect to spin and
valley indices, which are taken into account simply by multiplying
the number of states by 4. Spin degeneracy is usually justified by
the fact that typical Zeeman electronic level splitting induced by the
laboratory magnetic field is indeed very small. However, when
magnetic field is applied parallel to the plane of graphene, orbital motion and
Landau quantization are irrelevant and it is the lifting of spin and
valley degeneracies that becomes important \cite{Aleiner_PRB2007}. Such situation is in fact of immediate interest for possible spintronic applications, which employ the spin degree of freedom of Dirac fermions in graphene. With recent measurements showing spin coherence scale in graphene exceeding 1$\mu$m \cite{Tombros_Nature2007}, there is a growing recognition of the potential of this approach \cite{Tombros_Nature2007,Cho_APL2007,Hill_IEEE2006}. Moreover, it can be envisioned that spin-dependent splitting of the electronic levels in graphene could be induced by an effective magnetic field resulting from magnetic proximity effect in graphene in contact with a ferro/antiferromagnetic substrate \cite{patent,Haugen_2007}. Such ``exchange'' field acts only in the spin sector and can be much stronger than magnetic fields available in the laboratory, inducing level splitting sufficient for room-temperature device applications.

Here we consider transport of chiral Dirac fermions in graphene in the presence of such an in-plane magnetic field and their passage through a boundary between two regions with different field orientations. The latter might be associated with a domain wall in the magnetic layer of graphene-magnet (GM) heterostructure, and presents a basic element for spintronic applications.

\section{Effect of parallel magnetic field}

When external magnetic field is applied to graphene, its parallel component acts only on spin degree of freedom, while the perpendicular component couples both to spin and to the orbital motion as described by Eq.~\eqref{DiracEq}. Hence, the action of the parallel field is equivalent to band splitting by the ``exchange field'' arising from magnetic proximity effect induced by the magnetic substrate. Such proximity-induced field, regardless of its direction, does not couple to the orbital motion.

Point-like Fermi surface makes electronic properties of graphene extremely sensitive to external potentials. Experiments show that application of a moderate gate voltage results in an appearance of finite charge carrier density, which is proportional to the magnitude of applied electric field. The type of these induced carriers, revealed by the sign of the Hall effect, depends on the polarity of the gate voltage \cite{GeimNovoselov_NatMat2007,Novoselov_Nature2005}. This can be easily visualized by considering the electronic spectrum in graphene shown in Fig. 1(a), where the Fermi level is shifted by applying the gate voltage, thus inducing a circular Fermi-surface of particle or hole type, depending on the polarity.

External magnetic field $H$ splits electronic band structure in graphene according to spin. Chemical potential for one spin polarization is increased by the amount equal to Zeeman energy $g \mu_BH/2$, while for the other it is decreased by the same amount, Fig.~\ref{f:bands}(b) ($g \approx 2$ is the spectroscopic Lande factor for electrons in graphene, $\mu_B$ is Bohr magneton). As a result, there appear identical circular Fermi surfaces, of particle type for spin antiparallel to magnetic field, and of hole type for spin parallel to it. The radius of these Fermi surfaces, $\hbar k_H$, is proportional to the magnetic field,
\begin{equation}
\label{kH}
 \hbar k_H = g \mu_B H/ (2 v_F) \, .
\end{equation}
The difference in filling of the two spin states results in small Pauli paramagnetic moment
and total charge carrier density at the Fermi level,
\begin{equation}
\label{n_H}
 n(H) = (g \mu_BH)^2/(2\pi \hbar^2 v_F^2).
\end{equation}
In order to achieve non-negligible carrier densities, magnetic fields yielding Zeeman splitting of hundreds of Kelvins or more are required. While such fields can not be produced by solenoids, they might be induced by magnetic proximity effect in graphene-magnet multilayers \cite{patent,Haugen_2007}.

\section{The effective Hamiltonian}

In this section we review the low-energy Hamiltonian describing
electrons in graphene in the presence of Zeeman field. Recall that the Fermi surface of undoped graphene contains two non-eqiuvalent points, $K$ and $K'$, giving rise to a valley degeneracy. At each of these points the wave-function is a pseudo-spinor in the two-dimensional space of $A$ and $B$ sublattices and a spinor in the spin angular momentum space. We use Pauli matrices ${\tau}_{x,y,z}$, and ${\sigma}_{x,y,z}$ to refer to the sublattice pseudo-spin and the ``usual'' spin, respectively. With these notations, the effective Hamiltonian takes the form
\beq \label{Hamiltonian} \hat{H} = v_F \mathbf{p} \cdot \btau + {\bf
B}(x)\bsigma \, , \eeq
where ${\bf B} =g \mu_B \mathbf{H}(x)/2$, and spatially varying magnetic field $\mathbf{H}(x)$ couples to spin degree of freedom. As it was discussed above, we envision that this magnetic field can be induced by the proximity effect in graphene due to the super-exchange interaction with the magnetic layer contacting the graphene sheet \cite{patent,Haugen_2007}. As a consequence, magnetic field considered in Eq.~\eqref{Hamiltonian}, irrespective of its direction, acts only on spin degree of freedom of quasiparticles in graphene.
For the case of spatially homogeneous magnetization of graphene-magnet heterostructure, the spin component directed along the effective field is conserved. Hence, spin-polarized carriers maintain their polarization. Such structure can be used to transport spin-polarized currents. The basic element allowing manipulation of spin-polarized currents in graphene-magnet heterostructures is the region where magnetic field changes its direction, namely the domain wall (DW). In the next section we analyze transmission of the spin-polarized carriers through the domain wall.
\begin{figure}[h]
\includegraphics[width=0.9\columnwidth]{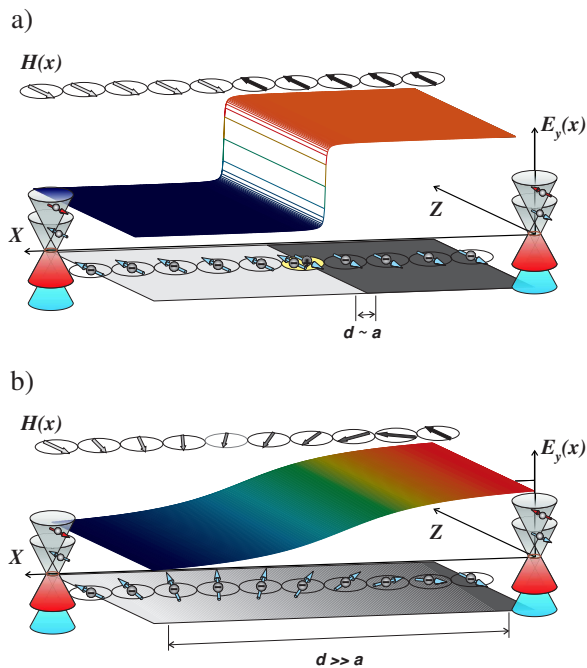}
\caption{Two regimes of spin-polarized electron transport through a domain wall between the regions with opposite spin polarizations in magnetized graphene. (a) Passage through narrow ($\lesssim 10$~ nm) domain wall is mediated by Klein tunneling of electron-hole pairs. (b) Transport through thick ($\gtrsim 100$~nm) domain wall allows electron's magnetic moment to follow the magnetic field direction.
}\label{f:domain}
\end{figure}

\section{Passage of spin-polarized Dirac fermions through a domain
wall}

In order to understand transport properties of an inhomogeneously magnetized graphene heterostructure, we analyze ballistic passage of spin-polarized carrier through the lateral domain wall separating two regions of opposite magnetization, Fig. \ref{f:domain}. The transmission through the domain wall is characterized by the amplitudes of spin flip and non-spin-flip processes. These amplitudes determine spin-polarized transport through DW and their knowledge is important for devising spintronics applications of the considered heterostructure.

For definiteness, let us consider electrons in the presence of the magnetic field $\mathbf{B}$ pointing along and opposite to $z$-axis on different sides of the domain wall located in the stripe region $x_0 < x < x_0 + L$. We further assume it to rotate uniformly within the DW (see Fig. \ref{f:domain}).
Specifically, we represent the magnetic field $\mathbf{B} = B \mathbf{n}(x)$ with the unit vector
\beq \label{DW-rotate}\mathbf{n}(x) = \left\{ \begin{array}{ll}
                    - \mathbf{n}_z & x < x_0 \\
                     -\mathbf{n}_z \cos\vartheta(x)  + \mathbf{n}_y \sin\vartheta(x)  & x_0 \leq x \leq x_0 + L \\                 \mathbf{n}_z & x > x_0 + L \\
                    \end{array}
                    \right.
                     \eeq
describing its rotation within the DW. In Eq.~\eqref{DW-rotate}, $\mathbf{n}_{i}$ stands for the unit vector pointing in the $i$-th direction, and the rotation angle is taken to be linear in the lateral coordinate $x$, $\vartheta(x) = \pi (x- x_0)/ L$.

We begin with the qualitative discussion of the carrier passage in the case of normal incidence. Due to the conservation of (pseudo-spin) chirality in the Klein tunneling phenomenon \cite{Novoselov_Nature2005} the backscattering is absent in this case. The carrier passes the DW during the passage time $t_P = L/v_F$, where $L$ is the characteristic width of the DW. Within the DW region the spin of the carrier experiences the time dependent torque and undergoes the Larmor precession. The parameter controlling the spin dynamics is the ratio of the passage time to the spin precession period,
\begin{equation}\label{eta}
\eta = 2 B L / \pi v_F \, .
\end{equation}
For thin DW, $\eta \ll 1$, and in the absence of the intervalley scttering,  $\lambda_{dB}/L \ll 1$, where $\lambda_{dB}$ is the carrier's de Broglie wavelength, the spin-flip probability is small, corresponding to a small precession angle. In the opposite limit of thick DW, $\eta \gg 1$, the spin follows the varying Zeeman field inside the DW adiabatically and the non-spin-flip probability is small. As a result, the carrier preserves its alignment with the field and reverses its polarization upon passing through the DW.
Although the scattering for the arbitrary angle is complicated by the finite backscattering amplitude, the basic physical picture presented above still holds and allows us to construct a scattering theory in the general case.

To solve the scattering problem we construct the scattering state,
\begin{align}\label{Sc-State}
\psi(x,z)\!=\!\begin{cases} \psi^{i}_{\alpha}(x,z) + r_{\alpha
\alpha' }\psi^r_{\alpha'}(x,z),\!&\!
x<x_0 \\
\,  t_{\alpha \alpha' } \psi^t_{\alpha'}(x,z),\!&\! x>x_0 + L
\end{cases}
\end{align}
where $\psi^{s}_{\alpha}(x,z)$ with $s=i,r,t$ denotes incoming, reflected and transmitted waves, and  the subscript $\alpha = \pm$ refers to the spin up (down) polarizations, respectively. Due to the translational symmetry in the $z$-direction, the scattering state in Eq.~\eqref{Sc-State} is the eigenstate of the $z$-component of the momentum and can be labeled by its eigenvalue $p_z$, making our problem effectively one-dimensional. We present wave functions entering Eq.~\eqref{Sc-State} in the following form
\begin{equation}\label{Sc-State-Spinors}
\psi^{s}_{\alpha}(x,z) = e^{i p_z z + i p^s_{x}(\alpha, p_z) x}
\varphi^s_{\alpha,p_z}\otimes \chi_{\alpha}\, , \quad s=i,r,l \, .
\end{equation}
Here $\chi_{\alpha}$ is the spin wavefunction, i. e. the spinor satisfying $\sigma_z \chi_{\alpha} = \alpha \chi_{\alpha}$, and $\varphi^s_{\alpha,p_z}$ is a pseudo-spinor in the sublattice space.

Capitalizing on the particle-hole symmetry of the problem, in what follows we only consider the case of incident quasiparticles with $E>0$. The incoming wave is an eigenstate of the Hamiltonian \eqref{Hamiltonian} for $x<x_0$. At a fixed energy, the majority (spin down) and minority (spin up) carriers have Fermi momenta $p_+ = (E+B)/v_F$ and $p_- = \left|\Delta E \right|/v_F$, respectively. Here $\Delta E = E - B$ can be both positive and negative, the latter case corresponds to the hole-like quasiparticles. The kinematic constraint for the incoming spin up (down) electrons reads $\left| p_z \right| < p_{\mp}$. Introducing the notation
\begin{subequations}\label{Notations}
\begin{equation}
u_{\pm}(p_z) = \sqrt{ ( 1 + p_z / p_{\pm} )/ 2}
\end{equation}
\begin{equation}
v_{\pm}(p_z) = \sqrt{ ( 1 - p_z / p_{\pm} )/ 2}
\end{equation}
\end{subequations}
for the pseudo-spinor $\varphi_{p_z}=\left( \cos \gamma(p_z)/2, \sin \gamma(p_z)/2 \right)^T$ forming angle $\gamma(p_z)$ with $z$-axis, we write the pseudo-spinor of the incoming electron as
\begin{subequations}\label{State-Spinors}
\begin{align}\label{I-State}
\varphi^i_{-,p_z} & = \left[ \begin{array}{ll} u_{+}(p_z) \\
v_{+}(p_z) \end{array} \right] \notag \\
\varphi^i_{+,p_z} & =
\theta_{\Delta E}
\left[ \begin{array}{ll} u_{-}(p_z) \\
v_{-}(p_z) \end{array} \right]
+%
\theta_{-\Delta E}
\left[ \begin{array}{ll} v_{-}(p_z) \\
u_{-}(p_z) \end{array} \right] \, ,
\end{align}
where $\theta_{\Delta E}$ and  $\theta_{- \Delta E}$ are step functions distinguishing cases of particle- and hole-like carriers, respectively.
In a similar fashion we write for the reflected wave
\begin{align}\label{R-State}
\varphi^r_{+,p_z}  & =
\theta_{\Delta E}
\left[ \begin{array}{ll} u_{-}(p_z) \\
-v_{-}(p_z) \end{array} \right]
+%
\theta_{-\Delta E}
\left[ \begin{array}{ll} v_{-}^*(p_z) \\
-u_{-}^*(p_z) \end{array} \right]
\notag \\
\varphi^r_{-,p_z}  & =
\theta_{\Delta E}
\left[ \begin{array}{ll} u_{+}(p_z) \\
-v_{+}(p_z) \end{array} \right]
+%
\theta_{-\Delta E}
\left[ \begin{array}{ll} u_{+}^*(p_z) \\
-v_{+}^*(p_z) \end{array} \right]\, .
\end{align}
The square roots in Eq.~\eqref{Notations} are defined as having positive imaginary part for negative argument. This choice, together with the sign and conjugation convention in Eq.~\eqref{R-State}, insures that for $ p_- < \left|p_z \right| < p_+ $ the wave function of the minority (spin up) carriers decays exponentially away from the DW, namely, it is an evanescent wave. The transmitted waves are given by
\begin{align}\label{T-State}
\varphi^t_{+,p_z}  & =
\left[ \begin{array}{ll} u_{+}(p_z) \\
v_{+}(p_z) \end{array} \right],
\notag \\
\varphi^t_{-,p_z}  & =
\theta_{\Delta E}
\left[ \begin{array}{ll} u_{-}(p_z) \\
v_{-}(p_z) \end{array} \right]
+%
\theta_{-\Delta E}
\left[ \begin{array}{ll} v_{-}^*(p_z) \\
u_{-}^*(p_z) \end{array} \right] .
\end{align}
\end{subequations}
Equations \eqref{State-Spinors} when substituted in \eqref{Sc-State-Spinors} give the explicit expressions for the incoming, reflected and transmitted spinors in the most general scattering state of Eq.~\eqref{Sc-State}.

In order to find the transmission and reflection amplitudes, the
transfer matrix $\widehat{T}$ matching the wave function at the two
boundaries of the DW
\begin{align}\label{match}
\psi(x_0) = \widehat{T}\psi(x_0 + L) \,
\end{align}
has to be found. To this end, we solve the Dirac equation inside the
wall,
\begin{multline}\label{inside}
\left( - i v_F \tau_x \partial_x + B\sigma_z \cos \vartheta(x)-
B\sigma_y \sin \vartheta(x) \right)\psi \\
=(E - v_F p_z \tau_z) \psi \, ,
\end{multline}
with the initial condition specifying the wave function at $x=x_0$.
The equation \eqref{inside} is formally equivalent to Rabi problem
of spin coupled to the oscillating magnetic field \cite{Rabi} with
coordinate $x$ playing the role of the time. The field is turned
on at the ``time'' $x = x_0$ and turned off at ``time'' $x=x_0 + L
$. Exploiting this analogy we solve Eq.~\eqref{inside}  by the
transformation to the rotating reference frame,
\begin{align}\label{frame-transf}
\psi(x) = \exp\left(-i \vartheta(x)\frac{\sigma_x}{2}\right)
\tilde{\psi}(x)
\end{align}
such that the field seen by the transformed spin is stationary.
Substitution of Eq.~\eqref{frame-transf} into Eq.~\eqref{inside}
gives
\begin{multline}\label{rot-frame}
\left( - i v_F \tau_x \partial_x - \frac{\pi v_F}{ 2 L } \tau_x
\sigma_x + B\sigma_z \right)\tilde{\psi}(x) \\
= (E - v_F p_z \tau_z) \tilde{\psi}(x)\, .
\end{multline}
We notice that static magnetic field now appears effectively as an operator in the pseudo-spin space.
We can rewrite Eq.~\eqref{rot-frame} in the form
\begin{equation}\label{rot-frame-1}
i \partial_{\zeta} \tilde{\psi} = \frac{ \pi }{ 2 } \hat{ A }
\tilde{\psi}(\zeta) \, ,
\end{equation}
where $\zeta = (x - x_0)/L$. The four-by-four matrix on the right
hand side of Eq.~\eqref{rot-frame-1} reads
\begin{equation}\label{A-matrix}
\hat{A} = -\sigma_x + \eta \tau_x \sigma_z - \eta \tau_x ( \epsilon - \epsilon_z
\tau_z)\,
\end{equation}
with dimensionless energies $\epsilon = E/B$ and $\epsilon_z = v_F p_z / B$ and the parameter $\eta$  defined in Eq.~\eqref{eta}.
The formal solution of equation \eqref{rot-frame-1} is
\begin{equation}\label{rot-frame-solve}
\tilde{\psi}(x_0 + L) = \exp\left( - i \frac{ \pi }{ 2 }
\hat{A}\right) \tilde{\psi}(x_0) \, .
\end{equation}
Combining equations \eqref{match}, \eqref{frame-transf} and
\eqref{rot-frame-solve} we obtain
\begin{equation}\label{W-matrix}
\widehat{T} = \exp\left( i \frac{ \pi }{ 2 } \hat{A}\right) i
\sigma_x\, .
\end{equation}
The exponentiation in Eq.~\eqref{W-matrix} can be performed
explicitly as follows
\begin{multline}\label{A-expon}
\exp\left( i \frac{ \pi }{ 2 } \hat{A}\right) =
 \sum_{\pm} \hat{P}_{\pm}  \\
\times
 \left[\cos\left(\frac{\pi}{
2 }\sqrt{c \pm \lambda }\right) - i \hat{A}\frac{ \sin\left( \frac{
\pi }{ 2 }\sqrt{ c \pm \lambda  } \right) }{ \sqrt{ c \pm \lambda }
} \right]\, ,
\end{multline}
where notations
\begin{align}
c & =  1 + \eta^2 + \eta^2(\epsilon^2 - \epsilon_z^2) \, , \\
\lambda & = 2 \eta \sqrt{\epsilon^2(1 + \eta^2) - \epsilon_z^2 } \notag
\end{align}
have been introduced, and
\begin{align}\label{P}
\hat{P}_{\pm} = \frac{\lambda \pm \hat{M} }{2 \lambda}
\end{align}
are projection operators onto the subspaces of the eigenvalues $\pm
\lambda$ of the matrix
\begin{align}\label{M}
\hat{M} = \hat{A}^2 - \frac{1 }{ 4} \mathrm{Tr}\hat{A}^2\, , \quad
\frac{1 }{ 4} \mathrm{Tr}\hat{A}^2 = c\, .
\end{align}
Equations \eqref{W-matrix}, \eqref{A-expon}, \eqref{P}, and
\eqref{M} give the transfer matrix $\widehat{T}$ introduced in
equation~\eqref{match} as a third degree polynomial in matrix
$\hat{A}$. For an arbitrary angle of incidence the transmission and
reflection coefficients are found by imposing the matching condition
in Eq.~\eqref{match} on the scattering state of
Eq.~\eqref{Sc-State}. The resulting system of four linear equations
determining four coefficients $r_{\alpha\alpha'}$,
$t_{\alpha\alpha'}$ can be easily solved. For the incoming (spin up)
minority carriers the probabilities of the passage with and without
spin flip are given by
\begin{align}\label{Prob}
T_{-+} &= \left| t_{-+} \right|^2 \notag \\
T_{++} &= \left| t_{++} \right|^2
\sqrt{ \frac{ 1 - (p_z/p_+)^2 }{ 1 - (p_z/p_-)^2 } } \, .
\end{align}

\begin{figure}[!th]
\includegraphics[width=1.0\linewidth]{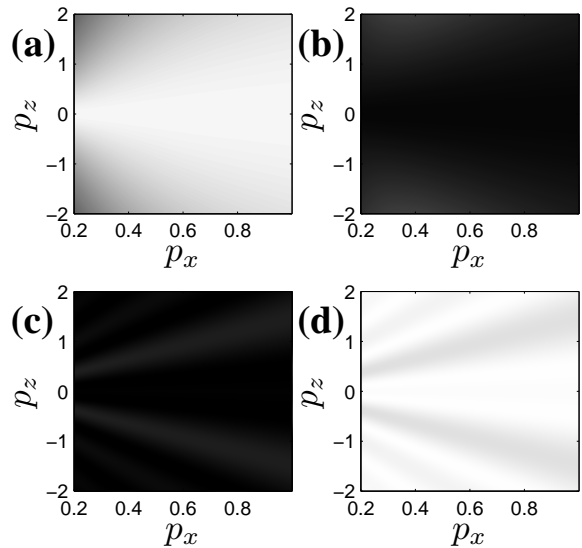}
\caption{Transmission probabilities for the incoming minority (spin up) carriers for two different DW thicknesses, $L$. The probability is shown on a linear gray scale from 0 (dark) to 1 (white). The parameters used are $B = 0.05$ and $v_F = 1$. Panels present the probability of transmission
(a) without the spin flip, $L = 5$ ;
(b) with the spin flip, $L = 5$;
(c) without the spin flip, $L = 50$;
(d) with the spin flip, $L = 50$.
The parameter $\eta$ defined by Eq.~\eqref{eta} is $\eta = 0.16$ for
the thin DW, (a), (b), and $\eta = 1.6$ for the thick DW, (c), (d).
}\label{fig:Transmission-Min}
\end{figure}

\begin{figure}[!h]
\includegraphics[width=1.0\linewidth]{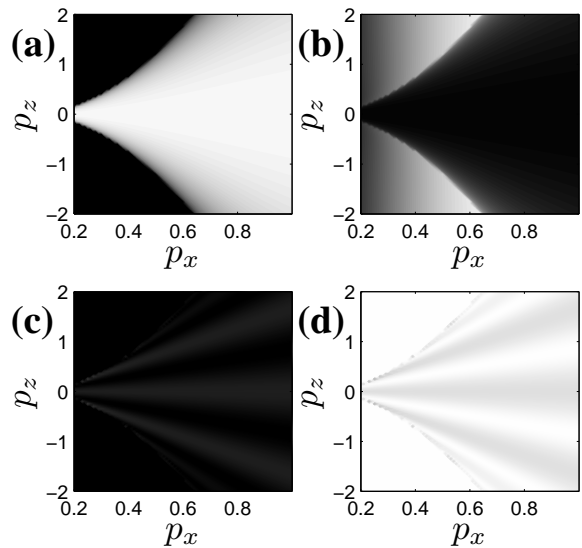}
\caption{Transmission probabilities for the incoming majority (spin down) carriers for the set of parameters used in Fig.~\ref{fig:Transmission-Min}
%
%
%
%
%
%
}\label{fig:Transmission-Maj}
\end{figure}

Although the solution of linear matching equations is straightforward, the final expressions are somewhat cumbersome and below we present the results graphically. The transmission probabilities for the incoming minority (spin up) and majority (spin down) species, respectively, are shown in Figs.~\ref{fig:Transmission-Min} and \ref{fig:Transmission-Maj}.
In both figures, panels (b) and (d) present the transmission with the spin flip, and panels (a) and (c) the transmission without the spin flip. Panels in the upper row show the case of the thin DW, $L = 5, \eta \ll 1$, and panels in the lower row the case of the thick DW,  $L = 50, \eta \gtrsim 1$.


\begin{figure}[!t]
\includegraphics[width=0.75\linewidth]{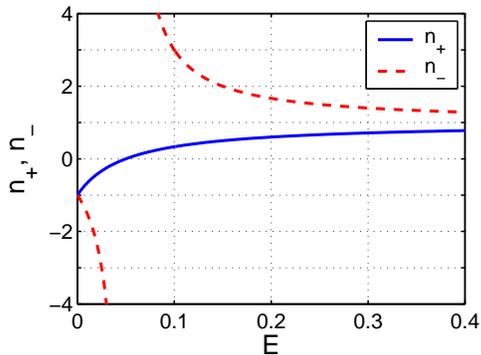}
\caption{Ratio of the refraction indices on the two sides of DW for the majority (-) and the minority (+) incident particles.
}\label{fig:Snell}
\end{figure}

Unlike the case of the normal incidence, for the incidence at an arbitrary angle the chirality is not conserved, leading to a finite  backscattering probability. Hence, particles passing the DW experience spin-dependent reflection and refraction. The Snell's law relating the angles of propagation of the incoming and the outgoing particles to the refraction indices of the two media reads
\begin{equation}\label{Snell}
 \frac{ n^{\pm}(x< x_0) }{n^{\pm}(x > x_0 + L ) }=\frac{ \sin \theta^{\pm}(x > x_0 + L ) }{\sin \theta^{\pm}(x< x_0)} \, .
\end{equation}
The conservation of the $z$-component of the momentum gives for the ratio of the refractive indexes,
\begin{equation}\label{Snell-ratio}
\frac{n^{\pm}(x< x_0)}{ n^{\pm}(x > x_0 + L ) } =
\frac{ E \mp B }{E \pm B }\, .
\end{equation}
The behavior of this ratio for two spin polarizations as a function of energy is shown in Fig.~\ref{fig:Snell}.

It follows from Eq.~\eqref{Snell-ratio} that the DW has a  different
effect on the spin minority and the spin majority carriers. The
minority carriers passing the DW experience the increase of the
optical density, while the majority carriers experience the decrease
of it. An interesting regime occurs when the energy of incoming
particles $E < B$. In this case the ratio in Eq.~\eqref{Snell-ratio}
becomes negative and results in the spin dependent Veselago lens
effect, similar to that discussed in
Ref.~\onlinecite{Cheianov-Science2008}.

The above spin-optics arguments are useful in understanding the results shown in Figs.~\ref{fig:Transmission-Min} and ~\ref{fig:Transmission-Maj}.
For the incidence at a shallow angle the probability of non-spin-flip passage is suppressed as particle trajectory even in the narrow DW   becomes long. The difference between the incoming minority and majority species in Figs.~\ref{fig:Transmission-Min}(a,b) and \ref{fig:Transmission-Maj}(a,b) results from the different refraction coefficients for the two species imposed by the kinematical constraints. The refraction coefficient ratio for minority carriers is $n_+ < 1$, and their trajectories bend so that the path inside the DW  shortens. For the majority carriers, on the other hand, $n_- < 1$, and the trajectory bending leads to the longer path inside the DW. Therefore, the effect of magnetic field inside the DW and the probability of transition without the spin flip are enhanced  for the minority carriers, Fig.~\ref{fig:Transmission-Min}(a), and suppressed for the majority carriers, Fig.~\ref{fig:Transmission-Maj}(a).

The Fabry-P\'{e}rot pattern of transmission seen in Figs.~\ref{fig:Transmission-Min}(d) and \ref{fig:Transmission-Maj}(d) is a consequence of an interference of multiple reflections inside the thick DW.

\subsection{Normal Incidence}
Our results are substantially simplified in the case of normal incidence, when the momentum component $p_z$ vanishes. With $\epsilon_z = 0$ the equation \eqref{A-expon} reduces to
\begin{widetext}
\begin{align}\label{A-expon-n}
\exp\!\left( i \frac{ \pi }{ 2 } \hat{A}\right) = 
 \sum_{\pm}\!\Bigg\{\!\left(\!\frac{ 1 }{ 2 } \pm \frac{ \sigma_x \tau_x - \eta \sigma_z }{ 2 \sqrt{ 1 + \eta^2 }
 }\!\right)
 \cos\!\frac{ \pi }{ 2 }\left( \sqrt{1 + \eta^2 } \pm \epsilon \eta \right) 
%
 + \frac{ i }{ 2  }\left(\!\frac{\sigma_x - \eta \sigma_z
\tau_x }{ \sqrt{ 1 + \eta^2 }} \pm \tau_x \!\right)\sin\!\frac{ \pi
}{ 2 }\left(\!\sqrt{1 + \eta^2 } \pm \epsilon \eta \!\right)
\Bigg\}\, .
\end{align}
\end{widetext}
The last equation gives for the transfer matrix
\begin{align}\label{W-n}
\widehat{T} =  i \sigma_x \cos\!\frac{ \pi }{ 2 }\!\sqrt{ 1\!+\!
\eta^2} - \frac{ 1\!+\!i\eta \tau_x \sigma_y }{ \sqrt{ 1\!+\!\eta^2}
}\sin\!\frac{ \pi }{ 2 }\!\sqrt{ 1\!+\!\eta^2 } \, ,
\end{align}
where the overall phase factor $e^{i \pi \epsilon \eta / 2} $ has
been omitted. In the present section we focus on the transmission
probabilities. It has to be stressed however that the phase of the
transition amplitude is also of interest, especially if the
magnetization vector $\mathbf{B}$ completes one or more rotation
circles inside the DW. Under the conditions of adiabatic spin
transfer, this phase is geometric~\cite{Berry1984}, see the
discussion of geometric Berry phase for the DW passage in
App.~\ref{App:geom}.

In the case of normal incidence the chirality is a good quantum
number, $[\widehat{T},\tau_x]_- = 0$. This ensures the absence of
backscattering (Klein tunneling phenomenon). For the incoming
particles with $E>0$  we have $\tau_x = +1$.  Therefore, dynamics in
the case of normal incidence occurs in the spin sector only. The
particle traveling inside the DW experiences the action of magnetic
field rotating with the frequency $\omega = \pi v_F /L$. The
probability of passing the DW without the spin flip is given by the
diagonal element of the transfer matrix
\begin{align}\label{P-no-flip}
T_{++} = \left(\frac{ \pi}{ \Omega t_P }\right)^2 \sin^2
\left(\Omega t_P / 2 \right) \, .
\end{align}
Here, $\Omega t_P = \pi \sqrt{ 1 + \eta^2 }$ is the rotation angle accumulated by the spin precessing at the Rabi frequency, $\Omega = \sqrt{ (\pi v_F/L)^2 + (2 B)^2 }$, during the passage time $t_P = L/v_F$.
It follows from Eq.~\eqref{P-no-flip} that the polarization of the
impinging particle is not influenced by the DW in the case of the
thin wall, $\eta \ll 1$. In the opposite limit of the thick DW, $\eta
\gg 1$, electron spin adjusts adiabatically following the direction
of the magnetic field slowly varying inside the DW.

Our results for the case of normal incidence are in agreement with
Ref.~\onlinecite{Newton1948}, where neutron polarization change in
the course of the passage through the ferromagnetic domain wall is
analyzed. We note that in the non-relativistic case
\cite{Newton1948} the absence of backscattering is an approximation,
which is valid for neutrons with high enough energy. This
approximation fails for low energy particles, as inside the DW
particles experience a force which results from the magnetic field
gradient inside the DW. In the present, relativistic, case however the
carriers with arbitrarily low energy can not be reflected by the DW
because of the chirality conservation. Therefore, in the
relativistic case the thick domain wall is $100 \%$ efficient in
flipping spins of particles incident at $90^\circ$.

Until now we have discussed the case of the DW with well-defined, abrupt boundaries, where the region of magnetic field variation is limited to a finite interval [see Eq.~\eqref{DW-rotate}].
While in most cases this is a reasonable description (in particular, for patterned structures), in some experimental realizations of spintronic devices the boundaries of the DW may be smooth and not well defined.
To clarify the significance of the above distinction in the DW
structure we consider the DW with the magnetic field direction
$\mathbf{n}=-\mathbf{n}_z \cos\vartheta(x)  + \mathbf{n}_y
\sin\vartheta(x)$ with the angle $\vartheta(x)$  following the
Rosen-Zener profile, $\partial_x \vartheta(x) =  (\pi / L) /
\cosh(\pi x / L)$. In this case the non-spin-flip transmission amplitude can be found exactly \cite{RosenZener},
\begin{equation} \label{RZ}
T_{++} = \mathrm{sech}^2\left( \frac{\pi \eta}{ 2 } \right) \, .
\end{equation}
It follows from comparison of Eq.~\eqref{P-no-flip} and
Eq.~\eqref{RZ} that the DW with smooth boundaries polarizes the
incoming carriers even more efficiently than the DW with abrupt
boundaries. Hence, we argue that both for thin DW, $\eta \ll 1$, and thick DW, $\eta \gtrsim 1$, cases our conclusions are valid for the DW of an arbitrary shape.
\section{Conductance in the ballistic transport regime}
In the high quality graphene devices the mean free path is comparable
to the characteristic sample size.
Under such conditions the transport through DW structures is ballistic. The conductance can be obtained within the Landauer approach. In spintronic devices we consider the spin selective transport, as they manipulate  the currents of the electrons of different polarizations independently.
Two terminal conductance is given by the sum of the
transmission probabilities over all active conductance channels.
Both majority and minority spin channels give rise to currents of
carriers of both polarizations. We introduce the spin dependent
conductance $G_{\alpha\beta}$ to denote the contribution of incoming carriers with spin $\beta$  in the source channels
to the current of carriers with spin $\alpha$ in the
drain channels,
\begin{equation}\label{Landauer}
G_{\alpha\beta} = \frac{ 2 e^2 }{ h } W
\int_{-k_{\alpha}}^{+k_{\alpha}} \frac{d p_z}{ 2 \pi } T_{\alpha
\beta}(p_z)\, ,
\end{equation}
where we have taken into account the 2-fold valley degeneracy.
The above definition is meaningful due to the conservation of $z$-component of spin away from the DW.

\begin{figure}[!th]
\begin{center}
\includegraphics[width=1.1\linewidth]{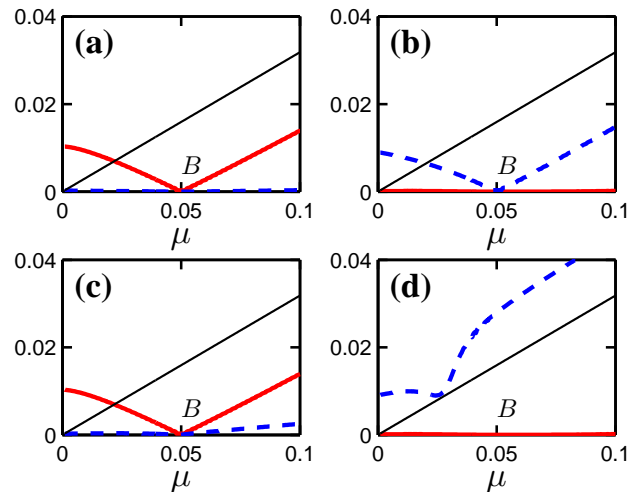}
\caption{(Color online) The partial conductances
$G_{\alpha\beta}/(2e^2 W/h)$ per unit sample width, $W$, as a
function of the chemical potential, $\mu$, for DW of two different
lengths $L$ and spin up $(\beta=+)$ and spin down $(\beta=-)$
incoming carriers, $B=0.05$. (a) $L = 5$, $\beta = +$, (b) $L = 50$,
$\beta = +$, (c) $L = 5$, $\beta = -$ and (d) $L = 50$, $\beta = -$.
In all panels the off diagonal conductances with $\alpha \neq \beta$
corresponding to a spin flip processes are shown by dashed (blue)
line, the diagonal conductances are shown by solid (red) line. The
thin (black) line shows the conductance in the absence of the DW per
one spin and one valley as a reference line. }
\label{fig:Conductance}
\end{center}
\end{figure}

Partial conductances $G_{\alpha\beta}/(2e^2W/h)$ obtained from
Eq.~\eqref{Landauer} are plotted in Fig.~\ref{fig:Conductance} as a
function of the chemical potential for two thicknesses of the DW,
$L=5$, (a,c) and $L=50$, (b,d).
The partial conductance for the positive spin polarization in the
source, $\beta = +$, is shown in the upper row,
Fig.~\ref{fig:Conductance}(a,b). That for $\beta = -$ is shown in
the bottom row, Fig.~\ref{fig:Conductance}(c,d). Solid lines
corresponds to the spin flip processes, namely $\alpha \neq \beta$
and dashed lines represent the diagonal conductances with $\alpha =
\beta$.

The common feature of the curves shown in Fig. \ref{fig:Conductance}
is the conductance growth when the chemical potential $\mu$ exceeds
half of the spin splitting, $B$. This is clearly due to the
increase of the number of conducting channels. Secondly, the increased thickness of the DW
stimulates spin flip processes. This is a consequence of the
adiabatic transfer of the spin inside the thick DW
discussed in Sec. IV. In addition, the only conductance not vanishing
at the special point $\mu = B$ in the case of the thick
DW is $G_{+-}$, see Fig. \ref{fig:Conductance}(d), dashed line. This is easily understood from the following consideration. For the thin DW the spin is approximately conserved. For that reason, the spin minority
(majority) carriers have vanishingly small number of incoming
(outgoing) channels  at $\mu \approx B$, leading to a small conductance. In the case of the thick DW the spin majority carrier can remain spin majority carrier by adiabatically adjusting (reversing) its spin polarization. This yields finite conductance at $\mu \approx B$. The specifics of the point $\mu = B$ described above makes hetero-structures with the Dirac spectrum of quasi-particles promising candidates for spin manipulation of the currents in spintronics.
\section{ Summary and Discussion}
In the present paper we have analyzed the passage of spin-polarized Dirac charge carriers through the DW in graphene-magnet hetero-structure. We have calculated the transmission and the reflection probabilities as a function of the energy of the incoming particles, which is determined by the average chemical potential $\mu$ in the graphene sample, for the DW of different thickness. The knowledge of the transmission amplitudes has allowed us to calculate the conductances of different spin channels in two-terminal geometry. The spin-polarized transport depends crucially  on the thickness of the DW.
Below we discuss the main features of our results and their potential applications in graphene-based spintronic elements.

We have considered two limiting cases of thin and thick DW. In the case of thin DW the spin dynamics inside the DW only occurs for shallow incidence angles. Aside from this special case, the spin is approximately conserved and the transmission is governed entirely by the kinematics of the relativistic Dirac quasiparticles in graphene, establishing direct connection with the problem of Klein tunneling and chiral dynamics in 2D quantum electrodymanics. A similar problem has recently been considered in the context of $p-n$ junctions in graphene devices~\cite{Cheianov-Science2008,Katsnelson_NatMat2006,Katsnelson_SSC2007}.
In our case the DW presents a $p-n$ junction for the majority and a $n-p$ junction for the minority carriers. The non-spin-flip transmission for $E < B$ is allowed through the particle-hole transmutation -- the Klein tunneling phenomenon.

The transmission properties of the thin DW can be nicely understood in the context of spin-optics~\cite{Khodas2007}. As it follows from the ratios of the refraction indices in Fig.~\ref{fig:Snell}, the trajectories of the minority (spin up) carriers bend inward, while those of the majority carriers bend outward. This difference is most pronounced near $E = B$, where the refraction index changes in sign, becoming negative for both spin polarizations at $E < B$. Near this point the transmission of the majority carriers through narrow DW vanishes and the corresponding refraction index diverges. The index of refraction for the minority carriers, on the other hand, is close to 0, as they can only pass through the DW near the forward direction. Hence, there is a giant birefringence of carriers with different spin polarizations, which could be employed in spin-selective transport devices. At $E \approx B$, the spin majority carriers undergo the total internal reflection due to the kinematic constraint, which occurs in a wide angular interval corresponding to dark areas in Fig.~\ref{fig:Transmission-Maj}(a). In the regime close to the total reflection the length of the particle's trajectory even inside a narrow DW becomes increasingly large. Hence, the probability of the spin-flip transmission becomes significant, see Fig.~\ref{fig:Transmission-Maj}(b). In this regime the angular aperture of the spin minority carriers becomes small, and thin DW is an efficient spin polarizer.

In the case of the thick DW the passage is governed by the spin dynamics inside the domain wall and the conductance is controlled  by the adiabatic nature of the spin transport. Independent of their energy, both majority and minority carriers simply flip their spins, remaining in their respective channels, so that thick DW acts as a spin-flipper for spin-polarized currents. As in the case of the thin DW considered above, the effect of the DW on the transport is most pronounced when the chemical potential is tuned to near the spin splitting, $\mu \approx B$.
In this case the Fermi surface of the minority spin carriers in the region $x>x_0 + L$ shrinks to a point and the current is carried by the majority species. In contrast to the case of the thin DW, the majority carriers in the region $x<x_0$ are transformed to the majority carriers in the region $x>x_0 +L$ by adiabatically adjusting their spin. This gives finite conductance for $\mu \approx B$, which is off-diagonal in spin, see Fig.~\ref{fig:Conductance}(d). Thus, a spin-transistor action could be achieved by virtue of adjusting the chemical potential in a gated device with magnetic layer coupled to the graphene layer.
%

Finally, manipulating the spin polarization of electrons in graphene by the means of Zeeman band splitting such as discussed in this paper depends crucially on the strength of the polarizing magnetic field. In the case of the passage through a DW, this strength also determines the relevant physical thickness distinguishing the cases of thick and thin DW, Eq.~\eqref{eta}. In the case of the laboratory magnetic field of $H_B \approx 1T$ created by a solenoid, the Zeeman band splitting is $B \approx 0.1$~meV. The condition $\eta \approx 1$ requires extremely thick DW, $L^* \approx 20 \mu$m. Moreover, such field results in a negligible spin-splitting, which corresponds to a temperature of only $\approx 1$ K, and negligible charge carrier density, $n_H \sim 10^{5} $~cm$^{-2}$, Eq~\eqref{n_H}.

However in the case of a spin-dependent band splitting induced by the magnetic proximity effect in graphene-magnet hetero-structure the corresponding effective spin-polarizing field could be expected to be of the order of hundreds, or even a thousand Kelvins. Physically, it could be estimated from the characteristic ordering temperature of the magnetic layer, and corresponds to 10~meV to 100~meV spin-dependent band splitting. Then, the effective ``exchange'' field could be as large as $\sim 10^3$~T, resulting in $n_H \sim 10^{11}$~cm$^{-2}$ and $L^* \sim 200$~nm.

Is it reasonable to expect such large magnetic proximity effects in graphene-magnet heterostructures? Theoretical estimates for the graphene layer in contact with EuO, one of the few insulating Heisenberg ferromagnets, predict band splitting of $\sim 5$ meV \cite{Haugen_2007}. This agrees well with the Curie temperature $T_C = 69.3$~K for this material. For room-temperature magnetic materials, the splitting could be expected to be proportionally higher. For example, one can envision using NiO [111] film in contact with graphene. NiO is an antiferromagnet with the Neel temperature $T_N \approx 500$~K, where ferromagnetically aligned Ni layers alternate in the [111] direction.
In the artificial [111] nano-layer CoO/NiO and NiO/Fe$_3$O$_4$ superlattices the proximity effect-induced increases of the ordering temperature by hundreds of Kelvins have been observed \cite{BorchersErwin_PRL1993}. It does not seem unreasonable to extrapolate this effect to graphene-magnet heterostructures. Finally, the technology of manufacturing graphene-based heterostructures is developing quite rapidly. The growth of the atomically smooth epitaxial MgO film on graphene has recently been reported \cite{Wang_APL2008}, as were nonvolatile memory devices obtained by covering graphene with a ferroelectric layer \cite{Ozyilmaz}. Therefore, manufacturing graphene-magnet heterostructures in order to achieve manipulation with spin-polarized currents such as considered in this paper looks reasonable and promising approach to be attempted experimentally.

In closing, we would like to note an interesting analogy between the generation of spin-polarized electric current in graphene hetero-structures and the separation of electric charge of quarks in
strong external magnetic fields in the presence of the topological charge in the quark-gluon plasma produced in relativistic heavy ion collisions \cite{Kharzeev:2004ey}.
There is a recent evidence\cite{Voloshin:2008jx} for this effect from STAR Collaboration at RHIC (BNL).
\begin{acknowledgements}
Illuminating discussions with A. Tsvelik, T. Valla and Y. Bazaliy
are greatly appreciated. This work was supported under the Contract
No. DE-AC02-98CH10886 with the U. S. Department of Energy. M. K.
acknowledges support from the BNL LDRD Grant No. 08-002.
\end{acknowledgements}
\begin{appendix}
\section{Geometrical phase of carriers passing the
DW}\label{App:geom}
The purpose of this appendix is to illustrate the general concept of
geometrical phases~\cite{Berry1984} for the spin carriers passing
the DW described by the Hamiltonian of Eq.~\eqref{Hamiltonian}. We
consider the normal incidence case where the dynamics occurs in spin
sector only. Here we are interested in the limit of adiabatic spin
transfer realized in sufficiently thick DW. The direction of the
magnetic field is a parameter of the system changing slowly inside
the DW,
\begin{equation}\label{app:slow}
\mathbf{B} = \mathbf{B}(x) + B_x \mathbf{n}_x \, ,
\end{equation}
where  $\mathbf{B}(x) = B \mathbf{n}(x) $, and
\begin{equation} \label{app:DW-rotate}\mathbf{n}(x) = \left\{ \begin{array}{ll}
                     \mathbf{n}_z & x < x_0 \\
                     \mathbf{n}_z \cos\vartheta(x)  - \mathbf{n}_y \sin\vartheta(x)  & x_0 \leq x \leq x_0 + L \\                 \mathbf{n}_z & x > x_0 + L \\
                    \end{array}
                    \right. \, .
                     \end{equation}
In the last equation we assume $\vartheta(x) = \Delta \theta (x
-x_0) /L $. For $\Delta \theta = 2 \pi$, the magnetic field defined
by Eqs.~\eqref{app:slow} and \eqref{app:DW-rotate} acting on a
particle passing the DW completes a circle, sweeping the conical
surface in $\mathbf{B}$-space. It is identical for $x<x_0$ and
$x>x_0+L$. Therefore, the wave function adiabatically following the
instantaneous eigenstate can only acquire the phase factor after the
DW passage,
\begin{equation}\label{app:2}
\psi(x_0 +L) = e^{i \phi_d } e^ {i \delta}  \psi(x_0) \, .
\end{equation}
The first factor represents the dynamical phase due to the spin
precession in magnetic field, and the second factor is geometric
Berry phase, independent of the DW structure in the adiabatic limit.
Below we calculate the geometrical phase explicitly for the specific
model of DW specified by Eqs.~\eqref{app:slow} and
\eqref{app:DW-rotate}, and compare it with the known theoretical
results.
We calculate transmission amplitudes $t_{\pm}$ for two spinors
$\psi_{+(-)}$ which are exact eigenstates of the Hamiltonian for
$x<x_0$. The spin in these states is polarized parallel
(anti-parallel) to the magnetic field outside the DW, i.e.
\begin{equation}\label{app:In-spinor}
\left\langle \psi_{\pm} \left| \bsigma  \right| \psi_{\pm}
\right\rangle = \pm\frac{(\eta_x, 0 ,\eta ) }{ \sqrt{ \eta^2 +
\eta_x^2 }  }\, .
\end{equation}
In the last equation we use the notation
\begin{equation}\label{eta-x}
\eta_x = 2 B_x L/\pi v_F
\end{equation}
similar to Eq.~\eqref{eta}.
The transmission amplitudes are diagonal elements of the inverse
transfer matrix $\widehat{T}$ defined in Eq.~\eqref{match},
\begin{equation}\label{app:amplit}
t_{\pm} = \left\langle \psi_{\pm} \left| \widehat{T}^{-1}  \right|
\psi_{\pm} \right\rangle \, .
\end{equation}
The unitary transfer matrix $\widehat{T}$ is found following the
same approach as in Sec.~IV
\begin{align}\label{app:T}
\widehat{T} & = \hat{U} \exp\left( i \pi \sigma_x \right) \exp\left( - i E L /v_F \right) \, , \notag \\
\hat{U} &= \cos \frac{ \pi }{ 2 } \sqrt{  \eta^2\! + \!(\eta_x \!-\!
\Delta
\theta/ \pi )^2 } \notag \\
+ & i \frac{ (\eta_x\! - \! \Delta \theta/ \pi )\sigma_x  + \eta
\sigma_z }{\sqrt{  \eta^2\! + \!(\eta_x \!-\! \Delta \theta/ \pi )^2
}}
\sin\!\frac{ \pi }{ 2 } \sqrt{  \eta^2\! +\! (\eta_x \!-\! \Delta
\theta/ \pi )^2 }\, .
\end{align}
Substituting Eqs.~\eqref{app:T} and \eqref{app:In-spinor} in
Eq.~\eqref{app:amplit} we obtain
\begin{align}\label{app:amplit1}
t_{\pm} =& \exp\left[\mp i  \frac{ \pi }{ 2 } \sqrt{ \eta^2 +
\eta_x^2 } + i \frac{ E L }{ v_F} \right] \notag \\
& \times \exp\left[ \mp i \pi \left( 1 - \frac{\Delta \theta \eta_x}
{ 2 \pi \sqrt{ \eta^2 + \eta_x^2 } }\right)\right] \, ,
\end{align}
where we made an expansion
\begin{align}\label{app:expand}
\sqrt{  \eta^2\! +\! (\eta_x \!-\! \Delta \theta/ \pi )^2 } \approx
 \sqrt{ \eta^2 + \eta_x^2 }
 - \frac{\Delta \theta \eta_x}{  \pi \sqrt{ \eta^2 + \eta_x^2 } }
\end{align}
valid in the adiabatic limit, $\eta,\eta_x \gg 1$.
Comparing Eq.~\eqref{app:amplit1} with Eq.~\eqref{app:2} we identify
the first factor as corresponding to the dynamical phase
\begin{align}\label{app:dynam}
\phi_d  = \frac{ E \mp \sqrt{ B^2 + B_x^2 } }{ L/v_F } \notag \\
\end{align}
due to the orbital motion and precession in the magnetic field, and
the second factor as the geometrical Berry phase,
\begin{align}\label{app:Berry}
\delta = \pm \pi ( 1 - \cos \varphi_B )\, ,
\end{align}
where $\varphi_B$ is an opening angle of the cone swept by the
magnetic field in the DW.  Equivalently, the phase $\delta$ is half
of the solid angle subtended by the closed contour in the
$\mathbf{B}$-space at the degeneracy point $\mathbf{B} = 0$, in
agreement with the general theory. We notice that the geometrical
phase is of opposite sign for two spin orientations $\psi_{\pm}$.
The resulting deviation of the precession angle from the one
expected from Eq.~\eqref{app:dynam} has been observed experimentally
for neutrons passing the region of the spiral magnetic
field~\cite{NeutronsBook}. We argue that similar phenomenon could be
observed in graphene samples for carriers passing the DW with the
magnetic field rotation angle, $\Delta \theta = 2 \pi $.
\end{appendix}

\end{document}